\documentstyle[prl,aps,preprint,epsfig]{revtex}
\tightenlines
\begin{document}

\newcommand{\be}{\begin{equation}}
\newcommand{\ee}{\end{equation}}
\newcommand{\ba}{\begin{eqnarray}}
\newcommand{\ea}{\end{eqnarray}}
\newcommand{\bk}{\mbox{\boldmath $k$}}
\newcommand{\rgl}{\rangle}
\newcommand{\lgl}{\langle}
\newcommand{\x}{\mbox{\boldmath $x$}}
\newcommand{\k}{\mbox{\boldmath $k$}}
\newcommand{\p}{\mbox{\boldmath $p$}}
\newcommand{\r}{\mbox{\boldmath $r$}}
\newcommand{\bu}{\mbox{\boldmath $u$}}
\newcommand{\nablab}{\mbox{\boldmath $\nabla$}}
\newcommand{\e}{\varepsilon}
\newcommand{\de}{\partial}
\newcommand{\nn}{\nonumber \\}
\newcommand{\lm}{{\ell m}}
\newcommand{\lmd}{{\ell' m'}}
\newcommand{\sqtpi}{\sqrt{\frac{2}{\pi}}}
\newcommand{\rhat}{\hat{\r}}
\newcommand{\Ainfl}{A_{\mbox{\scriptsize inf}}}
\newcommand{\planck}{_{\mbox{\scriptsize pl}}}
\newcommand{\perms}{\mbox{perms}}
\def\simless{\mathbin{\lower 3pt\hbox
   {$\rlap{\raise 5pt\hbox{$\char'074$}}\mathchar"7218$}}}  

\def\simgreat{\mathbin{\lower 3pt\hbox  
   {$\rlap{\raise 5pt\hbox{$\char'076$}}\mathchar"7218$}}}   
\def \pom {{\hspace{ -0.1em}I\hspace{-0.2em}P}}
\def \GeV {{\rm GeV}}
\def \MeV { {\rm MeV}}
\def \mb {{\rm mb}}
\def \mub {{\rm \mu b}}

\title{
	{\it In Press Phys. Rev D}\\
	\hspace{1.in}\\
	Non--Gaussian Signatures in the Cosmic Background Radiation from Warm
Inflation}

\author{S. Gupta\thanks{email: sg@roe.ac.uk} \& 
	A. Berera\thanks{email: ab@ph.ed.ac.uk} \& 
	A. F. Heavens\thanks{email: afh@roe.ac.uk} \& 
	S. Matarrese\thanks{email: matarrese@pd.infn.it}}
 
\address{$^*$ $^\ddag$Institute for Astronomy, 
	Royal Observatory,
	Blackford Hill,
	Edinburgh, 
	EH9 3HJ, U.K.\\
	$^\dag$Dept of Physics \& Astronomy, 
	University of Edinburgh, 
	Edinburgh, 
	EH9 3JZ, U.K.\\
	$^\S$Dipartimento di Fisica `G. Galilei', 
	Universit\`a di Padova, and INFN,
	Sezione di Padova, 
	via Marzolo 8, 
	I-35131, 
	Padova, 
	ITALY}

\maketitle

\begin{abstract}
We calculate the bispectrum of the gravitational field fluctuations generated during warm inflation, where dissipation of the vacuum potential during inflation is the mechanism for structure formation. The bispectrum is non--zero because of the self--interaction of the scalar field. We compare the predictions with those of standard, or `supercooled', inflationary models, and consider the detectability of these levels of non--Gaussianity in the bispectrum of the cosmic microwave background. We find that the levels of non--Gaussianity for warm and supercooled inflation are comparable, and over--ridden by the contribution to the bispectrum due to other physical effects. We also conclude that the resulting bispectrum values will be undetectable in the cosmic microwave background for both the MAP and Planck Surveyor satellites.

\vspace{0.34cm}
\noindent
PACS number(s): 98.80.Cq, 98.70.Vc, 98.80.Es
\end{abstract}

\section{Introduction}
The most commonly adopted model of the early Universe pictures
the large scale structure present in the Universe today to be seeded by
small scale fluctuations in the matter distribution of
the primordial universe during an inflationary phase.
In order to realize an inflationary regime,  the generic
dynamical model is
based on a single scalar field, often termed the inflaton.
The homogeneous, zero-mode component of this field is
pictured to roll on a ultraflat potential, thereby sustaining
a large potential energy and negligible kinetic energy,
which are the necessary conditions for realizing inflation.
These models assume there is no radiation production during
the inflation period, with any radiation present prior to
inflation rapidly diluting away.  We refer to this picture
as supercooled inflation, describing specifically
the thermodynamic state of the universe during inflation.
In this picture, the supercooled inflation phase is separated 
from the radiation--dominated regime
%of the Friedman--Robertson--Walker picture of standard cosmology
by a brief reheating period, in which vast amounts of radiation
are rapidly produced.
%terminating inflation and in 
%which all the radiation needed to initiate the radiation dominated regime
%is generated very quickly.

In supercooled inflation the initial seeds of density perturbations
result from quantum fluctuations
of the inflaton field.  To a good approximation, the fluctuations
have a Gaussian distribution and produce a nearly scale--invariant
spectrum. A perfect Gaussian distribution would imply that the
density perturbations have no connected correlations higher
than the 2--point correlation function in real and Fourier space.
However, the self--interaction of the inflaton field is known to produce
non-zero, but extremely small, non--Gaussian effects, and there have been predictions calculated of these effects \cite{ganguietal} and their detectability in the distribution of the cosmic microwave background (CMB) fluctuations 
\cite{verdetal,wangkam}. Other structure formation scenarios, such as the class
of defect models, and multiple--field inflation models, generally give larger 
deviations from Gaussianity.

There are many ways of testing the Gaussian hypothesis, such as 
the genus and Euler-Poincar\'e statistic
\cite{Coles88,Gott90,Luo94b,Smoot94}, studies of tensor modes in the CMB \cite{CCT94}, excursion set
properties \cite{Barreiro98,Barreiro2000}, peak
statistics \cite{BE87,Kogut95,Kogut96,Barreiro97} and wavelet analyses (e.g. \cite{MHL00,AF99,FA99}). Then there are the set of higher--order 
correlation functions, such as the three-point function
(e.g. \cite{Hinshaw94,FRS93,LS93,ganguietal}), the bispectrum \cite{Luo94,heavens,ferreira}, and the trispectrum \cite{Kunzetal}. A significantly 
non--Gaussian 
signal in the CMB sky, measured from the data of the Cosmic Background 
Explorer (COBE) Differential Microwave
Radiometer (DMR) satellite instrument, launched in 1989, has been claimed 
\cite{ferreira}, but doubts have been cast on this as a significant primordial signal, and more recent papers using this data \cite{Kunzetal,banday,kometal} and, on a smaller angular scale, the MAXIMA balloon 
experiment data of 1998 \cite{santosetal,wuetal} find results largely 
consistent with Gaussianity.

Gangui {\it et al.} \cite{ganguietal} calculated predictions of the bispectrum for
several variants of the supercooled inflationary scenario.
In the limit of no instrument or sampling noise, the minimum variance
on CMB data is the cosmic variance \cite{smoot_isot} -- arising from fact that our Universe could only be one of a Gaussian ensemble. Gangui {\it et al.} \cite{ganguietal} found the values for skewness for the bulk of their single field models to be considerably smaller in
magnitude than that resulting from cosmic variance. The skewness is, however, 
only a single statistic which is part of a much wider class -- the three--point correlation function. There is the prospect of doing much better by using the whole class or, equivalently, its harmonic counterpart, the bispectrum.

COBE had an angular resolution of $7^{\circ}$. For COBE it has been possible 
to analyse all the modes of
the bispectrum up to the resolution limit \cite{komsperg,ferreira}. The 
Planck Surveyor satellite is planned for launch
in 2008. It will have an angular resolution of down to 5 arcminutes. 
The Microwave 
Anisotropy Probe (MAP) is operating now, with a resolution of $12.6'$. Both of 
these experiments allow vast numbers of bispectrum modes to be analysed in principle, and it is interesting to see if either could distinguish warm inflation 
from supercooled inflation on the basis of the bispectrum.

In this paper we predict the form of the fluctuations and quantify the
non--Gaussianity, using the bispectrum as a measure, for
warm inflation dynamics.
Warm inflation \cite{wi} differs from the standard,
supercooled picture of inflationary cosmology in that the process of
radiation production becomes an important constituent of the
theory.  In particular, radiation production occurs concurrently
with inflationary expansion and the presence of this radition
influences the seeds of density perturbations.

The warm inflation picture of inflationary dynamics is a 
comprehensive set of possible interactions between fields 
during inflation.   In this picture no a priori assumptions are made 
about multi-field interactions, thus particle
production, during the inflationary epoch.  
As such, the warm inflation picture makes explicit that
the thermodynamic state of the universe during inflation 
is a dynamical question.
In particular, the supercooled inflation emerges as
one limiting case in which the interactions are negligible.
More commonly dissipation is possible in warm inflation with a
resulting density of radiation present during inflation \cite{wi,bf2,ab2}.
Here both strong \cite{wi,bf2,ab2} and weak \cite{bf2,weakd,fl}
dissipative regimes have been examined which offer several
variants to the basic picture.

A variety of warm inflation models have been developed at a 
phenomenological level \cite{ab2,weakd,phenomen,taylber}. 
From this it has generally been understood
that warm inflation can solve the basic cosmological puzzles
of horizon, flatness and density fluctuations.  However, up
to now no study has been made on the degree of non--Gaussianity
typically emerging in warm inflation models and it is important
to quantify such effects.
  
In this paper, a general methodology
is developed for computing non--Gaussian effects in warm inflation
scalar field models.  The statistic we use to quantify the resulting non--Gaussianity in the cosmic microwave background predicted for the case of warm inflation models is the bispectrum. This formalism is then applied to the $\lambda \phi^4$, $\lambda \phi^3$ and $m^2 \phi^2$ models for the strong dissipative regime of warm inflation in the interest of discovering whether these effects are comparable, or indeed distinguishable, from the predictions of supercooled inflation.

\section{Warm Inflation Dynamics}

The equation of motion for the
zero mode of the scalar inflaton field, $\phi$, in general has the form
\be
	\ddot{\phi} + 3H\dot{\phi} + \Gamma\dot{\phi} + V'(\phi) =0,
\label{eq:infeom}
\ee
where the overdots represent time derivatives, $H=\frac{\dot{R}}{R}$
is the Hubble parameter, and $R(t)$ is the cosmic expansion factor.
The necessary condition for
inflation is domination of the inflaton
potential energy over all other
energy components in the universe.  This is achieved by
requiring the inflaton potential to have sizable magnitude and
be very flat.  The flatness of the potential allows slow roll motion
of $\phi$, so that inflaton kinetic energy becomes neglible
with the effect that the ${\ddot \phi}$ term can be dropped from
the equation of motion.  In supercooled inflation
dissipative effects, which in Eq.(\ref{eq:infeom}) are symbolically
expressed through the term $\Gamma {\dot \phi}$,
are assumed negligible during
the inflation period and only emerge during
the subsequent reheating period.  On the other hand,
the basic observation of warm inflation is that inflationary
conditions remain energetically possible even in
the presence of a sizable radiation component.  Thus
in warm inflation dissipation remains active during inflation,
with the simplest representation of these effects
being the form in Eq.(\ref{eq:infeom}).  The resulting evolution
equation for the inflaton in warm inflation therefore is
\be
	\frac{d\phi}{dt}=-\frac{1}{3H + \Gamma}\frac{dV(\phi)}{d\phi}
\label{eq:wi}
\ee
where similar to supercooled inflation, the slow-roll condition
is required,
$(3H+\Gamma)|\dot{\phi}|\gg |\ddot{\phi}|$.

The presence of dissipation implies throughout the inflation
period radiation is produced from conversion of
vacuum energy.  In the context of Friedmann cosmology,
the stress energy conservation equation
consisting of a radiation, $\rho_r$, and a vacuum energy component, $\rho_v$, is
\be
	\dot{\rho_r}(t) = -4\rho_r(t) H -  \dot{\rho}_v(t) .
\label{eq:strss_en}
\ee
If there were there no dissipation, then
${\dot \rho}_v =0$ and the radiation
component would be rapidly red--shifted away as
$\rho_r\sim e^{-4Ht}$. However with dissipation, radiation
is being produced continuously from
conversion of scalar field vacuum energy.
For dissipation of the form in Eq.(\ref{eq:infeom}),
$-{\dot \rho}_v = \Gamma {\dot \phi}^2$.

Note that in general the conversion of vacuum energy into radiation
will result in some type of reaction back upon the scalar
field, with the $\Gamma {\dot \phi}$ term being
the simplest phenomenological possibility.  Attempts to obtain
first principles scalar field evolution equations
for warm inflation have obtained dissipative
% Fix this
effects which in general are temporally nonlocal although
limits also have been obtained in which the effects are
of the local form in Eq.(\ref{eq:infeom})\cite{wi,bgr,beramos}.

The production of radiation during inflation in general will
influence the seeds of density fluctuations.  In particular,
if the temperature during inflation
is bigger than the Hubble parameter, $T > H$, then
structure formation can be significantly affected by the thermal
component.  In a preliminary work leading to the development
of the warm inflation scenario \cite{bf2}, it was shown that
a tiny dissipative component $\Gamma \simgreat 10^{-5}H$
already is adequate to realize $T > H$.  Subsequently
the phenomenology of both weak $\Gamma < H$ \cite{bf2,weakd}
and strong $\Gamma \geq H$ \cite{ab2,bgr,bgr2,abadiab} 
dissipative regimes have been
studied for warm inflation.  More attention has been given
to the strong dissipative regime.  This primarily
is because the main focus of first principles quantum field
theory studies of warm inflation \cite{bgr,bgr2,yoko,abadiab,beramos,bk,bk2}
have been in this regime, since it is the more difficult
of the two, and once this regime is understood the weak
dissipative regime easily would follow.
In this paper, expressions will be obtained which apply to the strong dissipative regime, such that Eq.(\ref{eq:wi}) reduces to $\frac{d\phi}{dt}=-\frac{V'}{\Gamma}$, and during inflation $\dot{\rho}_r\sim 0$, $4\rho_r H\sim -\dot{\rho}_v$.

In order to treat the fluctuations of the inflaton field
$\delta \phi({\bf x},t)$, it is assumed that they are
small and the full inflaton field is expressed as
$\phi({\bf x},t) = \phi_0(t) + \delta \phi({\bf x},t)$
with $\phi_0$ being the homogeneous `background' field, $\delta \phi({\bf x},t) \ll \phi_0(t)$.
The equation of motion for the fluctuations of the inflaton field
can be obtained by imposing a near--thermal--equilibrium, Markovian
approximation, which therefore implies the fluctuation--dissipation
theorem is applicable.  From this
the equation of motion for the full inflaton field emerges as
\be
	\frac{d\phi(\x,t)}{dt}=\frac{1}{\Gamma}
\left[e^{-2Ht}\nabla^2
\phi(\x,t) - V'(\phi(\x,t)) + \eta(\x,t) \right].
\label{eq:wi_pert}
\ee
Implementing the fluctuation--dissipation theorem immediately determines the properties of the noise. With respect to physical coordinates and in momentum space,
these properties are
\begin{equation}
\langle \eta \rangle=0
\label{eq:noise1}
\end{equation}
\begin{equation}
\langle \eta({\bk},t) \eta({\bk}',t') \rangle=
2\Gamma T (2\pi)^3 \delta^{(3)}({\bk}- {\bk}')
\delta(t-t')
\label{eq:noise2}
\end{equation}

\section{The Statistics of Warm--Inflation Perturbations}

% put bisp bit here
\subsection{The Predictions and Properties of Gaussian Fields}

Single field inflation models broadly predict Gaussian primordial density
fluctuations. Multiple field inflation models may lead to 
a non--Gaussian distribution (e.g. chi--squared). 
When second--order effects are taken into account, however, 
there are corrections to
these general predictions. There is a resulting non--Gaussian signal in
the CMB, the magnitude of which varies depending upon the
self-interaction of the inflaton field \cite{FRS93,luoschramm,ganguietal}.

The statistical properties of a Gaussian field with mean zero are
fully contained in its power spectrum or its two--point correlation
function in real space. Higher order correlations can be expanded in
terms of these quantities, so correlation
functions of even order of the distribution can be written as products
of two--point functions, and correlations of odd--order can be written in
terms of products of two--point functions and the expectation value of the
field. As there are no connected correlations over the two--point, all
odd order correlation functions of a multivariate Gaussian
distribution with zero mean are equal to zero. For a non--Gaussian field, the higher--order connected correlation function can be non--zero. 

The three--point correlation function of the density perturbation distribution in Fourier space, otherwise known as the {\it bispectrum}
\cite{Luo94,heavens,ferreira,ganguietal} is the quantity we
have chosen to evaluate as our measure of the non--Gaussianity
generated by warm inflation. This quantity translates to the harmonic bispectrum of the CMB. 

\subsection{The Warm Inflation Bispectrum}
%Pasted upto here

We will begin by expanding Eq.(\ref{eq:wi_pert}) in order to obtain the  
evolution equations up to second order in the fluctuations
$\delta \phi({\bf x},t)=\delta \phi_1({\bf x},t) + \delta \phi_2({\bf x},t)$, where $\delta \phi_1={\mathcal O}(\delta \phi)$ and $\delta \phi_2={\mathcal O}(\delta \phi^2)$. All calculations that follow will be in momentum space, and with respect to
the physical versus comoving momenta, where recall
\be
{\bk}_{\mbox{\scriptsize phys}} ={\bk}_{\mbox{\scriptsize com}} e^{-Ht}.
\label{eq:phys_k}
\ee
Hereafter for physical momenta, the notation will have
no subscripts ${\bk}_{\mbox{\scriptsize phys}} \equiv {\bk}$,
and magnitudes will be denoted without boldfacing as
$k \equiv |{\bk}|$. 
The choice of physical coordinates arises since we are interested
in the evolution of the inflaton mode while they are sub--horizon scale.
During this time, the dominant effect on the modes is from 
the high temperature heat bath, and these effects are more conveniently
analyzed in physical coordinates.  When evaluating the
equation of motion for the fluctuations, the time
dependence of the physical modes will be treated adiabatically
with respect to the characteristic macroscopic time scale,
the Hubble time $\sim 1/H$.  Thus the evolution equations
to be written for the inflaton modes
will only be valid over a time interval $\sim 1/H$
and a complete solution over longer time for a given mode can be 
obtained by piecewise construction.  Thus the equations
of motion of the first and second order fluctuations
over a time period $\sim 1/H$ are

\ba
	\frac{d}{dt}(\delta\phi_1(\k,t))= \frac{1}{\Gamma}[-k^2
\delta\phi_1(\k,t) - V''(\phi_o(t))\delta\phi_1(\k,t) \nn + \eta(\k,t) ]
\label{eq:wisol1}
\ea
\ba
	\frac{d}{dt}(\delta\phi_2(\k,t))= \frac{1}{\Gamma}[-k^2
\delta\phi_2(\k,t) - V''(\phi_o(t))\delta\phi_2(\k,t) \nn - \frac{1}{2}V'''(\phi_o(t))\delta\phi_1(\k,t)^2 ],
\label{eq:wisol2}
\ea
where $\delta \phi({\bk},t) = \delta \phi_1({\bk},t) +
\delta \phi_2({\bk},t)$ is the inflaton mode with physical
momentum ${\bk}$ at cosmological -- corresponding to the
homogeneous background -- time $t$.
The properties of the noise are given in Eqs.(\ref{eq:noise1},\ref{eq:noise2}).
We also assume here that the evolution of 
scalar field fluctuations can be studied in a particular gauge 
where metric perturbations can be neglected compared with those of 
the scalar field itself. The latter assumption, originally made 
in Ref.\cite{FRS93}, allows one to focus on the computation of those 
non--Gaussian 
features which are directly produced by non--linearities (i.e. 
self--interactions) of the scalar field itself, rather than by their  
backreaction on the underlying geometry. As we will see, this 
approach will give rise to a level of non-Gaussianity which is 
comparable to that obtained in Ref.\cite{ganguietal}, where 
backreaction effects were simply modelled through a local modification 
of the Hubble expansion rate during inflation.  

Dividing cosmic time into successive time intervals of
order $1/H$, $t_n - t_{n-1} = 1/H$, the solutions of
Eqs. ({\ref{eq:wisol1}) and (\ref{eq:wisol2}) for $t_{n-1} < t < t_n$ are respectively 
\ba
\lefteqn{\delta\phi_1(\k,t)=} \hspace{5mm} \nn
&&A(k,t-t_{n-1})\int^t_{t_{n-1}}dt'\frac{\eta(\k,t')}{\Gamma}
A(k,t'-t_{n-1})^{-1} \nn &&+
A(k,t-t_{n-1})\delta\phi_1(\k e^{-H(t_n-t_{n-1})},t_{n-1})
\label{eq:sol1}
\ea
\ba
\lefteqn{\delta\phi_2(\k,t)=} \hspace{5mm} \nn
&&A(k,t-t_{n-1})\int^t_{t_{n-1}}dt' \nn
&& B(t')\left[\int\frac{dp^3}{(2\pi)^3}
\delta\phi_1(\p,t')\delta\phi_1(\k-\p,t')\right]
A(k,t'-t_{n-1})^{-1} \nn 
&& + 
A(k,t-t_{n-1})\delta\phi_2(\k e^{-H(t_n-t_{n-1})},t_{n-1}) \;,
\label{eq:sol2}
\ea
where
\ba
A(k,t)=\exp \left[-\int^{t}_{t_o}\left( \frac{k^2}{\Gamma}+
\frac{V''(\phi_o(t'))}{\Gamma}\right)dt' \right]
\label{eq:fn_A}
\ea
\ba
	B(t)=-\frac{V'''(\phi_o(t))}{\Gamma}.
\label{eq:fn_B}
\ea

In both solutions Eqs. (\ref{eq:sol1}) and (\ref{eq:sol2}), the second
term on the RHS are ``memory'' terms that reflect the 
state of the given mode at the beginning of the given time interval.
The relevance of these memory terms leads to the important 
concept of freeze--out \cite{abadiab}.  By definition
of freeze--out, for $|{\bk}| \stackrel{>}{\sim} k_F$
the memory terms damp away within a Hubble time and 
for $|{\bk}| \stackrel{<}{\sim} k_F$ they do not.  To
quantify this criterion, the freeze--out momentum $k_F$
is defined by the condition
\be
	\frac{k^2+V''(\phi_o)}{H\Gamma}>1
\label{eq:therm_kF}
\ee
In general for warm inflation $V''(\phi_o)<\Gamma H$,
 so the above
condition can be simplified to $k_F=\sqrt{\Gamma H}$.
In supercooled inflation the freeze--out wavenumber would
correspond to the Hubble scale, as the quantum fluctuation becomes
classical on horizon exit. For warm inflation the fluctuations freeze
in {\it before} horizon exit.

Freeze--out implies noteworthy features about the solutions.
When $k > k_F$, since the memory terms are negligible, $\delta \phi_1$,
$\delta \phi_2$ primarily are determined by the state of
the environment nearby in time.  Then, when $k < k_F$, 
$\delta \phi_1$, $\delta \phi_2$ are determined dominantly 
by their state at time of freeze--out.  These two facts imply
the time--slicing approach we used to solve for
$\delta \phi_1$, $\delta \phi_2$ is well justified.

From the solutions Eqs. (\ref{eq:sol1}) and (\ref{eq:sol2}), we are interested
in computing the three-point correlation function of the inflaton
fluctuations at the largest observable scales, which
in particular cross the horizon in the interval
$\sim 50 - 60$ e-folds before the end of inflation.
During this time interval, 
we will compute the defining parameters of the three--point
correlation function, which
as will be seen can be expressed in terms of a 
bispectrum that well approximates the Sachs--Wolfe
regime, $l \stackrel{<}{\sim} 50$.  Thus
from the three-point function
$\langle \delta \phi({\bk}_1, t_{60}) \delta \phi({\bk}_2, t_{60})
\delta \phi({\bk}_3, t_{60}) \rangle$,
the amplitude and slope of the bispectrum are determined
at time $t \approx t_{60}$, 60 e-folds before the end of
inflation and for ${\bk}_1$, ${\bk}_2$, ${\bk}_3$
all all within a few e--folds of exiting the horizon.

The leading order contribution to this three-point correlation
function comes from two first-order and one second-order
fluctuation as 

\ba
\lefteqn{\langle\delta\phi(\k_1,t)\delta\phi(\k_2,t)\delta\phi(\k_3,t)\rangle=}
  \hspace{5mm} \nn
& &  A(k_3,t-t_{60}-1/H)\int^{t_{60}}_{t_{60-1/H}}A^{-1}(k_3,t'-t_{60}-1/H)B(t')\nn
& &  \left[\int\frac{dp^3}{(2\pi)^3}
\langle\delta\phi_1(\k_1,t_1)\delta\phi_1(\p,t')\rangle
\langle\delta\phi_1(\k_2,t_2)\delta\phi_1(\k_3-\p,t')\rangle\right]\nn
& & 
+ A(k_3,t-t_{60}-1/H) 
\langle\delta\phi_1(\k_1,t_{60})\delta\phi_1(\k_2,t_{60})
\nn
& & \delta\phi_2(\k_3e^{-1},t_{60}-1/H)\rangle
\nn 
&&  +(\k_1 \leftrightarrow \k_3))
+(\k_2 \leftrightarrow \k_3))
\label{eq:inf_bisp}
\ea
In this expression, since $B(t')$ is slowly
varying, it can be approximated as a constant.
Similarly $\delta \phi({\bk},t)$ can be fixed at its freeze--out
value.  The three-point function on the RHS arises from the memory
term of $\delta \phi_2$ in Eq. (\ref{eq:sol2}), since $k < k_F$.  In evaluating
this quantity, first note the coefficient in front can be
approximated as unity, $A(k, t_{n-1}-t_{n-2}) \approx 1$.
Furthermore, since all three momenta will overlap in
the freeze--out region, the three-point correlation
function at $t_{60} - 1/H$ is approximately the same
as at $t_{60}$, and this property will repeat itself for the time interval
\begin{equation}
\Delta t_F \equiv t_H - t_F \approx
\frac{1}{H} \ln(\frac{k_F}{H}),
\end{equation}
where $t_H$ represents the time at Hubble
crossing of the smallest of the three inflation perturbation modes, and $t_F$ represents the time when the last of the three
wavevectors thermalizes. Thus Eq. (\ref{eq:inf_bisp}) becomes
\ba
\lefteqn{\langle\delta\phi(\k_1,t)\delta\phi(\k_2,t)\delta\phi(\k_3,t)\rangle
\approx B(t_{60}) \Delta t_F} \hspace{5mm} \nn
&& \left[\int\frac{dp^3}{(2\pi)^3}\langle\delta\phi_1(\k_1,t_1)\delta\phi_1(\p,t')\rangle\langle\delta\phi_1(\k_2,t_2)\delta\phi_1(\k_3-\p,t')\rangle \right. \nn
&& \left. +(\k_1 \leftrightarrow \k_3)) + (\k_2 \leftrightarrow \k_3)
\right].
\label{eq:infbisp2}
\ea

\subsection{Estimating the Magnitude of the Non--Gaussianity}

It is possible to write a general expression for the bispectrum for
slow roll, single field, supercooled inflation models as well as for
the set of warm inflation models.
\ba
\lefteqn{\langle\Phi(\k_1)\Phi(\k_2)\Phi(\k_3)\rangle=}\hspace{5mm}\nn
&& \Ainfl (2\pi)^3\delta^3(\k_1+\k_2+\k_3) \left[P_\Phi(\k_1)P_\Phi(\k_2)+\perms\,\right]
\label{eq:bisp_a_infl}
\ea
The relation between the scalar field fluctuation and the
gravitational field has the simple form \cite{bard}

\be
\Phi(\k)=-\frac{3}{5}\frac{H}{\dot{\phi}}\delta \phi(\k),
\label{eq:bardeen}
\ee
thus $\Ainfl$ for a strongly dissipative warm inflation
regime is
\ba
\label{eq:ainfl_wi}
\Ainfl^{\mbox{\scriptsize warm}}=-\frac{10}{3}\left(\frac{\dot{\phi}}{H}\right)
\left[\frac{1}{H}\ln\left(\frac{k_F}{H}\right)\frac{V'''(\phi_o(t_F))}{\Gamma}\right] \;.
\ea

Comparative estimates for the non--Gaussianity of the cosmic
microwave background can then be calculated using $\Ainfl$ and the
shape of the inflationary potential.
To estimate the magnitude of $\Ainfl^{warm}$, consider the model
\be
V(\phi)=\frac{\lambda}{4!} \phi^4,
\label{eq:potential}
\ee
in the region $0< \phi < M$, where, fixing the origin
of time at $t=0$, initially $\phi(0)=M$. Here the self--coupling 
constant $\lambda$ is dimensionless. $M$ sets a basic scale
such as the Grand Unified scale $\sim
10^{14}$GeV, although the final answer for $\Ainfl^{warm}$
is independent of this scale.   The solution of the zero-mode
evolution equation Eq. (\ref{eq:wi}) for this case is
\be
\phi_0(t) = \frac{M}{\left(\frac{\lambda M^2}{3\Gamma}t +1\right)^{1/2}} .
\label{eq:phi}
\ee
In \cite{ab2} it is shown the number of e-folds of inflation $N_e$
for this quartic potential is
\ba
\label{eq:Ne}
N_e \approx \frac{1}{2}\left(1+\frac{48\pi\Gamma^2}
{3m\planck^2\lambda}\right)^{1/2} .
\ea
The fluctuations in the scalar field caused by thermal interactions
with the radiation field are \cite{abadiab}
\be
\delta\phi^2=\frac{k_F T}{2\pi^2} ,
\label{eq:delphi}
\ee
The temperature, $T$
can be calculated using the relation
$\rho_r=\left(g_\ast\pi^2/30\right)T^4$,
where $g_\ast$
is the number of relativistic fields $\sim 150$, 
taking into account the 
relation between the radiation energy density and the scalar field
potential presented in Eq. (\ref{eq:strss_en}).
The CMB amplitude is given by 
\be
\delta_H = \frac{2}{5}\frac{H}{\dot{\phi}}\delta \phi \;,
\label{eq:del_h} 
\ee
where from COBE data it is measured to be
$\delta_H = 1.94 \times 10^{-5}$ \cite{smoot_isot,bunlidwhit,bunnwhite}.

Setting $N_e=60$ and using 
Eqs. (\ref{eq:phi}), (\ref{eq:Ne}) and (\ref{eq:del_h}),
the value of $\lambda$ is found to be 
\be
\lambda=7.2\times10^{-15}
\ee
and this gives us all the quantities needed to evaluate
$\Ainfl$ for warm inflation, Eq. (\ref{eq:ainfl_wi})
\be
\Ainfl^{\mbox{\scriptsize warm}}= 7.44\times10^{-2} .
\label{awarm}
\ee The equivalent quantity to $\Ainfl$ for supercooled inflation
appears in Gangui {\it et al.} \cite{ganguietal} as
$\Phi_3\equiv\Ainfl^{\mbox{\scriptsize
supercooled}}=5.56\times10^{-2}$ for a quartic potential. A similar
quantity, $f_{NL}$, appears in \cite{komsperg} and a related quantity
in \cite{wangkam}, $A_{\mbox{\scriptsize infl}}$. They are related as
\be 
\Phi_3\equiv\Ainfl\equiv2f_{NL}\equiv A_{\mbox{\scriptsize infl}}
(2\pi)^6 \;.
\ee 
For this potential we see that the non--Gaussianity in the
curvature in warm inflation models is comparable to that in
supercooled inflation. Furthermore due to
the slow-roll behavior, for general models with local behavior of the
form $\phi^q$ with $2 < q \leq 4$, the value of $\Ainfl^{warm}$
will be within the same order of magnitude as
Eq. (\ref{awarm}). Note, however, that this is only the
contribution arising from non--Gaussianity in the inflaton field.  In
addition, there is a contribution at the last-scattering surface
arising from second-order gravitational perturbation theory.  Even at
this relatively early time, the nonlinear gravitational equations
induce a contribution $\Ainfl^{\mbox{\scriptsize 2nd order}}={\mathcal
O}(1)$, \cite{LS93,moller,mun,pynecarr,komsperg,molmat,spergold}, which is
therefore the dominant source of non--Gaussianity in both warm
inflation and supercooled inflation models.  The prospects of
measuring this are poor; the bispectrum of the microwave background
radiation is the most obvious possibility, and this has been
investigated by \cite{komsperg}. Note that this assumes
that the Hubble parameter is unchanged at horizon exit for all modes
considered.  The prospects are best for the Planck satellite
\cite{bersn}, as this measures modes up to $\ell \sim
2000$, but even with the optimistic assumption that all foreground
contaminants can be removed perfectly, the expected error on $f_{NL}$
is still 5. Since Silk damping causes the power to decline rapidly on
scales smaller than the Planck beam, there is little prospect of a
future, higher-resolution experiment being able to improve
significantly on this. Similarly, polarisation measurements are
unlikely to help, since most of the power is expected at $\ell \sim
100$, for which there are relatively few bispectrum coefficients.

\section{Conclusions}

We have explored the evolutionary behaviour of gravitational field
fluctuations generated by warm inflation, up to second order in the
scalar field perturbations, for the particular case of 
strong dissipation. We have
obtained predictions for the non--zero bispectrum of the gravitational
perturbations due to the self--interaction of the inflaton field, and
the resulting harmonic bispectrum for the Sachs--Wolfe region on the
cosmic microwave background. Eq.(\ref{eq:bisp_a_infl}) represents the
gravitational bispectrum in a generic form, with the value of $\Ainfl$
resulting from the theory of the fluctuation generation mechanism. The
CMB bispectrum can also be related to $\Ainfl$, provided that the
potential is sufficiently flat that the Hubble parameter etc are
approximately constant on horizon exit for all modes considered.  We
find that the inherently classical mechanism for the generation of
fluctuations in warm inflation, which is palpably different from the
corresponding mechanism for supercooled inflation, produce a level of
non--Gaussianity of approximately the same magnitude, $\Ainfl\sim
10^{-2}$.  This arises from non--Gaussianity in the inflaton field
itself.  The dominant contribution to the curvature bispectrum,
however, comes from general relativistic second--order perturbation
theory, which contributes an $\Ainfl \sim 1$, in both warm and
supercooled inflation.  Unfortunately this level of non--Gaussianity in
the CMB bispectrum appears to be unobservable, even with the Planck
satellite.  Conversely, any measure of primordial non--Gaussianity
significantly in excess of $f_{NL}=1$ would rule out both warm
inflation and supercooled inflation, although some alternatives such
as the curvaton model \cite{lythwands} or multi-field 
inflation models (e.g. \cite{bmr3} and refs. therein) could survive.

Acknowledgements:

SG and AB thank PPARC for financial support.

\vspace{0.5cm}


\begin{references}

%1
\bibitem{ganguietal} A. Gangui, F. Lucchin, S. Matarrese and
S. Mollerach, ApJ, {\bf 430}, 447 (1994)
%2
\bibitem{verdetal} L. Verde, L. Wang, A. F. Heavens and M. Kamionkowski, MNRAS,
{\bf 313}, 141 (2000)
%3
\bibitem{wangkam} L. Wang and M. Kamionkowski, Phys. Rev. {\bf D61},
063504 (2000)
%4
\bibitem{Coles88} P. Coles, MNRAS, {\bf 234}, 509 (1989)
%5
\bibitem{Gott90} J. R. Gott, C. Park, R. Juszkiewicz, W.E. Bies, D.P. Bennett, F.R. Bouchet and A. Stebbins, ApJ, {\bf 352}, 1 (1990)
%6
\bibitem{Luo94b} X. Luo, Phys. Rev {\bf D49}, 3810 (1994b)
%7
\bibitem{Smoot94} G. F. Smoot, L. Tenorio, A. Banday, A. Kogut, E. L. Wright, G. Hinshaw and C. L. Bennett, ApJ, {\bf 437}, 1 (1994)
%8
\bibitem{CCT94} D. Coulson, R.G. Crittenden and N. Turok, Phys. Rev. Lett., {\bf 73}, 2390 (1994)
%9
\bibitem{Barreiro98} R. Barreiro , J. Sanz, E. Martinez--Gonzales and J. Silk, MNRAS, {\bf 296}, 693 (1998)
%10
\bibitem{Barreiro2000} R. Barreiro, E. Martinez-Gonzales and J. Sanz, MNRAS, {\bf 322}, 411 (2001)
%11
\bibitem{BE87} J. R. Bond and G. P.  Efstathiou, MNRAS, {\bf 226}, 655 (1987)
%12
\bibitem{Kogut95} A. Kogut, A. Banday, C.L. Bennett, G. Hinshaw, P.M. Lubin 
and G.F. Smoot, ApJ {\bf 439}, L29 (1995)
%13
\bibitem{Kogut96} A. Kogut, A. Banday, C.L. Bennett, K.M. Gorski, G. Hinshaw, G.F. Smoot and E.L. Wright, ApJ {\bf 464}, L29 (1996)
%14
\bibitem{Barreiro97} R. Barreiro, J. Sanz, E. Mart\'{i}nez--Gonz\'{a}les and J. Silk, ApJ {\bf 478}, 1 (1997)
%15
\bibitem{MHL00} P. Mukherjee, M. P. Hobson and A. N. Lasenby, MNRAS,
{\bf 318}, 1157 (2000)
%16
\bibitem{AF99} N. Aghanim and O. Forni, A\&A, {\bf 347}, 409 (1999)
%17
\bibitem{FA99} O. Forni and N. Aghanim, A\&A Suppl., {\bf 137}, 553 (1999)
%18
\bibitem{Hinshaw94} G. Hinshaw, A. Kogut, K.M. Gorski, A.J. Banday, C.L. Bennett, C. Lineweaver, P. Lubin, G.F. Smoot and E.L. Wright, ApJ, {\bf 431}, 1 (1994)
%19
\bibitem{FRS93} T. Falk, R. Rangarajan and M. Srednicki, ApJLett, {\bf 403}, 1 (1993)
%20
\bibitem{LS93} X. Luo and D. Schramm, Phys. Rev. Lett., {\bf 71}, 1124 (1993)
%21
\bibitem{Luo94} X. Luo, ApJLett, {\bf 427}, 71 (1994a)
%22
\bibitem{heavens} A. F. Heavens, MNRAS, {\bf 299}, 805 (1998)
%23
\bibitem{ferreira} P. Ferreira, J. Magueijo and K. Gorski, ApJLett, {\bf 503}, 1 (1998)
%24
\bibitem{Kunzetal} M. Kunz, A. J. Banday, P. G. Castro, P. G. Ferreira and K. M. Górski, ApJLett, {\bf 563}, L99 (2001)
%24a
\bibitem{banday} A. J. Banday,  S. Zaroubi and K. M. Górski, ApJ, {\bf 533}, 575 (2000)
%24b
\bibitem{kometal} E. Komatsu, B. D. Wandelt, D. N. Spergel, A. J. Banday and K. M. Górski, ApJ, {\bf 566}, 19 (2002)
%25
\bibitem{santosetal}  M. G. Santos, A. Balbi, J. Borrill, P. G. Ferreira, S. Hanany, A. H. Jaffe, A. T. Lee, J. Magueijo, B. Rabii, P. L. Richards, G. F. Smoot, R. Stompor, C. D. Winant and J. H. P. Wu, astro--ph/0107588
%26
\bibitem{wuetal} J. H. P. Wu, A. Balbi, J. Borrill, P. G. Ferreira, S. Hanany, A. H. Jaffe, A. T. Lee, B. Rabii, P. L. Richards, G. F. Smoot, R. Stompor and C. D. Winant, Phys. Rev. Lett. {\bf 87}, 251303 (2001)
%27
\bibitem{smoot_isot} G. F. Smoot, C. L. Bennett, A. Kogut, J. Aymon, C. Backus, G. de Amici, K. Galuk, P. D. Jackson, P. Keegstra,
L. Rokke, L. Tenorio, S. Torres, S. Gulkis, M. G. Hauser,
M. A. Janssen, J. C. Mather, R. Weiss, D. T. Wilkinson, E. L. Wright,
N. W. Boggess, E. S. Cheng, T. Kelsall, P. Lubin, S. Meyer,
S. H. Moseley, T. L. Murdock, R. A. Shafer and R. F. Silverberg, ApJ,
{\bf 371L}, L1 (1991)
%28
\bibitem{komsperg} E. Komatsu and D. N. Spergel, Phys. Rev. {\bf D63},
063002 (2001) 
%29
\bibitem{wi} A. Berera,  Phys. Rev. Lett. {\bf 75}, 3218 (1995);
A. Berera, Phys. Rev. {\bf D54} (1996) 2519
%30
\bibitem{bf2} A. Berera and L. Z. Fang, Phys. Rev. Lett. {\bf 74}, 1912
(1995)
%31
\bibitem{ab2} A. Berera, Phys. Rev. {\bf D55}, 3346 (1997)
%32
\bibitem{weakd} H. P. de Oliveria and R. O. Ramos, 
Phys. Rev. D{\bf 57} 741 (1998);
H. P. de Oliveira and S. E. Jorás, 
Phys. Rev. D{\bf 64} 063513 (2001)
%33
\bibitem{fl} W. Lee and L. Z. Fang, J. Mod. Phys. {\bf D6}, 305 (1997)
%34
\bibitem{phenomen} W. Lee and  L.-Z. Fang, Phys. Rev. {\bf D59}, 083503 (1999);
J. M. F. Maia and J. A. S. Lima, Phys. Rev. {\bf D60},
101301 (1999);  M. Bellini, Class. Quant. Grav. {\bf 16}, 2393 (1999);
L. Chimento, {\it et. al.}, astro--ph/0201002
%35
\bibitem{taylber} A. N. Taylor and A. Berera, Phys. Rev. {\bf D62},
083517 (2000)
%36
\bibitem{bgr}A. Berera, M. Gleiser and R. O. Ramos, Phys. Rev. {\bf
D58} 123508 (1998)
%37
\bibitem{bgr2} A. Berera, M. Gleiser and R. O. Ramos,
Phys. Rev. Lett. {\bf 83} 264 (1999)
%38
\bibitem{abadiab} A. Berera, Nuclear Phys. B, {\bf 585}, 666 (2000)
%39
\bibitem{yoko} J. Yokoyama, A. Linde, Phys. Rev. {\bf D60}, 083509 (1999)
%40
\bibitem{beramos}A. Berera and R. O. Ramos, Phys. Rev. {\bf
D63} 103509 (2001)
%41
\bibitem{bk} A. Berera and T. W. Kephart, Phys. Lett. {\bf B456} (1999) 135.
%42
\bibitem{bk2} A. Berera and T. W. Kephart, 
Phys. Rev. Lett. {\bf 83} 1084 (1999)
%43
\bibitem{coulcritur} D. Coulson, R. G. Crittenden, N. G. Turok,
Phys. Rev. Lett., {\bf 73}, 2390 (1994)
%44
\bibitem{luoschramm} X. Luo, D. Schramm, Phys. Rev. Lett., {\bf 71},
1124 (1993)
%45
\bibitem{bard} J. M. Bardeen, Phys. Rev. {\bf D22}, 1882 (1980)
%46
\bibitem{bunlidwhit} E. F. Bunn, A. R. Liddle and M. White, 
Phys. Rev. {\bf D54},
5917 (1996)
%47
\bibitem{bunnwhite} E. F. Bunn and M. White, ApJ {\bf 450},
477 (1995)
%48
\bibitem{moller} S. Mollerach, A. Gangui, F. Lucchin and S. Matarrese, ApJ
{\bf 453}, 1 (1995)
%49
\bibitem{mun} D. Munshi, T. Souradeep and A. Starobinsky, ApJ {\bf 454} 
552 (1995)
%50
\bibitem{pynecarr} T. Pyne and S. M. Carroll, Phys. Rev. {\bf D53}, 2920 (1996)
%51
\bibitem{molmat} S. Mollerach and S. Matarrese, Phys. Rev. {\bf D56}, 4494 
(1997)
%52
\bibitem{spergold} D. N. Spergel and D. M. Goldberg, Phys. Rev. {\bf D59}, 
103001 (1999)
%53
\bibitem{bersn} M. Bersanelli, F.R. Bouchet, M. Griffin,
          J.M. Lamarre, N. Mandolesi,
          H.U. Norgaard-Nielsen, O. Pace, J. Polny,
          J.L. Puget, J. Tauber, N. Vittorio and
          S. Volont\'e, COBRAS/SAMBA Report on the Phase A study ESA D/SCI 
          (1996)
%54
\bibitem{lythwands} D. H. Lyth and D. Wands, Phys. Lett. {\bf B524}, 5 (2002)
%55
\bibitem{bmr3} N. Bartolo, S. Matarrese and A. Riotto, Phys. Rev. {\bf D65}, 103505
(2002)

%
\end{references}
\end{document}